# Synthesis of a new alkali metal-organic solvent intercalated iron selenide superconductor with Tc≈45K.

A. Krzton-Maziopa[1,*], E.V. Pomjakushina[1], V. Yu. Pomjakushin[2], F. von Rohr[3], A. Schilling[3], K. Conder[1]

[1]*Laboratory for Developments and Methods, Paul Scherrer Institute, 5232 Villigen PSI, Switzerland*

[2]*Laboratory for Neutron Scattering, Paul Scherrer Institute, 5232 Villigen PSI, Switzerland*

[3] *Physik-Institut der Universität Zürich, Wintherthurerstrasse 190, CH-8057 Zürich, Switzerland*

E-mail: ekaterina.pomjakushina@psi.ch

[*] *On leave from Warsaw University of Technology, 00-664 Warsaw, Poland*

**Abstract.** We report on a new iron selenide superconductor with a $T_C$ onset of 45K and the nominal composition $Li_x(C_5H_5N)_yFe_{2-z}Se_2$, synthesized via intercalation of dissolved alkaline metal in anhydrous pyridine at room temperature. This superconductor exhibits a broad transition, reaching zero resistance at 10K. Magnetization measurements reveal a superconducting shielding fraction of approximately 30%. Analogous phases intercalated with Na, K and Rb were also synthesized and characterized. The superconducting transition temperature of $Li_x(C_5H_5N)_yFe_{2-z}Se_2$ is clearly enhanced in comparison to the known superconductors $FeSe_{0.98}$ (Tc ~ 8K) and $A_xFe_{2-y}Se_2$ ($T_C$ ~ 27-32K) and is in close agreement with critical temperatures recently reported for $Li_x(NH_3)_yFe_{2-z}Se_2$. Post-annealing of intercalated material ($Li_x(C_5H_5N)_yFe_{2-z}Se_2$) at elevated temperatures drastically enlarges the *c*-parameter of the unit cell (~44%) and increases the amount of superconducting shielding fraction to nearly 100%. Our findings indicate a new synthesis road leading to possibly even higher critical temperatures in this class of materials by intercalation of organic compounds between Fe-Se layers.

## 1. Introduction

The discovery of superconductivity with $T_C$ around 30K in layered chalcogenides $A_xFe_{2-y}Se_2$ with A = K, Cs, Rb (122) with $ThCr_2Si_2$- structure type [1,2,3] evoked intensive studies concerning the improvement of their superconducting properties. Extended X-ray and neutron diffraction studies performed on those materials demonstrated a presence of a superstructure resulting from Fe vacancies ordering with a $\sqrt{5}\times\sqrt{5}\times1$ supercell below ~560K and strong antiferromagnetism with a magnetic transformation at high temperatures reaching $T_N$ ~ 500K. The value of the ordered magnetic moment of Fe amounts to 3.31 $\mu_B$ at 11K, which is the largest among all known iron based superconductors [4,5,6]. Recent μSR [7], STM [8] and Raman [9] studies showed a mutual relation between superconductivity and magnetism in these materials. It has been also noted that the evolution from a superconducting state to an insulating AFM state for 122 intercalated iron chalcogenides depends both on the Fe and alkali metal content. The ideal compositions (still under debate) [10,11,12] are difficult to control during synthesis by conventional high temperature methods.

A pronounced reversible phase separation revealed recently in 122 single crystals [13,14,15,16,17,18], as well as controversies regarding the origin of superconductivity and the exact stoichiometry of the superconducting phase [16,19,20] are still in the forefront of scientific activity. Further, investigations on alternative synthesis routes of superconducting 122 phases have been undertaken. Since alkali and alkaline-earth metals dissolve easily in liquid ammonia, such solutions were successfully used in the past for intercalation of layered chalcogenides [21,22]. Therefore its choice as the reaction environment for intercalation of layered iron chalcogenides seemed to be the most natural. The first positive report about superconducting iron selenides intercalated with alkali and alkaline-earth metals from liquid ammonia solutions [23] presented a variety of compounds with significantly enhanced $T_C$ ~40-46K. However, they possessed small superconducting volume fractions between 0.5 and 30% as derived from ZFC/FC DC magnetization. Shortly after that, lithium-ammonia co-intercalated compounds with nearly 80% shielding fractions were reported [24,25]. In the work of Burrard-Lucas *et al.* [24] both lithium amide and ammonia as intercalated guests were considered while Scheidt *et al.* [25] stress the role of $Li_x(NH_3)_y$ as the only spacer involved in the intercalation process. Despite some discrepancies regarding the exact nature of guest moieties, all the reports are consistent with respect to the value of the superconducting transition temperature and the crystal structure of the intercalated material. The crystal structure of $Li_xFe_2Se_2(NH_3)_y$ has been indexed as a body centered tetragonal with I4/mmm symmetry. It has been also noted that lattice parameters of $Li_xFe_2Se_2(NH_3)_y$ change slightly with the amount of the intercalated species, showing *a* = 3.8373(6) and *c* = 16.518(3) Å for $Li_{0.9}Fe_2Se_2(NH_3)_{0.5}$, whereas in the case of material containing twice more Li cations and the co-intercalated lithium amide $Li_{1.8}Fe_2Se_2(NH_3)[Li(NH_2)]_{0.5}$, *a* = 3.79607(8) and *c* = 16.9980(11) Å. In comparison to the 122-iron chalcogenides, the in-plane lattice parameter is slightly smaller but the c-parameter is dramatically expanded. Thus, these findings suggest further investigation on larger molecular spacers which can be co-intercalated together with alkali metal cations in the host Fe-Se matrix. Furthermore, in the case of iron chalcogenides intercalated from ammonia solutions, a gradual deintercalation caused by the loss of $NH_3$ was observed [24]. Therefore, application of less volatile species, i.e. heterocyclic amines being able to dissolve and then to coordinate lithium in the host matrix instead of ammonia, seems to be a good alternative. Aromatic amines, such as pyridine and its derivatives, have been widely used as moieties capable to intercalate layered dichalcogenides [26,27] and metal oxy- and nitride halides [28,29]. Anhydrous pyridine dissolves alkali metals producing highly reactive species of the general formula: $A(C_5H_5N)_2$. Depending on the molar ratio of the alkali metal to pyridine, the 4,4'-bipyrydyl radical anion can be also formed creating an ionic compound with the alkali metal cation [30]. We therefore supposed that both these compounds, as well as the free pyridine molecules can be successfully intercalated between Fe-Se host layers.

In the present work we report on the synthesis of a new hybrid superconductor $A_x(C_5H_5N)_yFe_{2-z}Se_2$ (A = Li) with organic spacers co-intercalated between Fe-Se layers together with alkali metal. Analogous phases with Na, K and Rb were also prepared and characterized. For A = Li, we obtain bulk superconductivity below Tc ≈ 40K. Additional treatment of the intercalated materials at elevated temperatures drastically increases the *c*-parameter of the unit cell and improves the superconducting properties.

## 2. Experimental

Alkali metal intercalated samples of the general formula $A_x(C_5H_5N)_yFe_{2-z}Se_2$ (A = Li, Na, K, Rb) were obtained via room temperature intercalation into the iron selenide matrix in pyridine solutions of the corresponding alkali metals. Iron selenide (FeSe) was synthesized from high purity (at least 99.99%, Alfa) powders of iron and selenium according to the procedure described elsewhere [31]. For the intercalation process an appropriate amount of a powdered FeSe precursor was placed into a container filled with 0.2M solution of pure alkali metal dissolved in anhydrous pyridine. The amount of FeSe taken for intercalation was calculated for the molar ratio 1:2 of alkali metal and precursor, respectively. The reaction was carried out at 40°C until the discoloration of the alkali metal solution. After synthesis the intercalated material was separated from the solution, washed repeatedly with fresh pyridine and dried up to a constant mass in inert atmosphere. All the work was performed in a He glove box to protect the powder from oxidation.

The phase purity of the samples was characterized by powder X-ray diffraction (XRD) using a D8 Advance Bruker AXS diffractometer with Cu K$\alpha$ radiation. For these measurements a low background airtight specimen holder was used. Rietveld refinements of the diffraction data were performed with the FullProf package [32].

Elemental composition of the obtained samples was studied using x-ray fluorescence spectroscopy (XRF, Orbis Micro-XRF Analyzer, EDAX). Elemental distribution maps for K, Na, Rb, Fe and Se were collected in vacuum by applying white X-ray radiation produced by an Rh-tube (35kV and 500μA) and performing the similar standardization as described before [10]. Please note that analysis of Li is not possible by this method.

The ac magnetic susceptibility measurements were performed with a Quantum Design PPMS magnetometer ($H_{ac}$ = 1Oe), the resistivity measurements were performed on a Quantum Design PPMS equipped with a resistivity option.

Thermogravimetric measurements have been performed on NETZSCH STA 449C analyzer equipped with PFEIFFER VACUUM ThermoStar mass spectrometer. All the experiments have been done in a stream of He, heating the samples up to 450 °C with a rate of 5 °C/min.

## 3. Results and discussion

Fig.1 shows the XRD pattern for the $K_x(C_5H_5N)_yFe_{2-z}Se_2$ sample. The material contains practically one phase with the $A_xFe_{2-y}Se_2$ (122) structure [16]. The only impurity detected in the sample is the FeSe parent compound with ≈4% mass fraction. The most intense peak from the FeSe phase is located at 2Θ=28.6°. Although the average structure of the main phase can be well described by the 122 vacancy disordered structure (I4/mmm space group), one can see the superstructure satellite shown in the inset of Fig 1. The intensity of the satellite is compatible with the √5x√5 vacancy ordered structure [5] (space group I4/m) with an almost empty (4d) position of Fe. In the refinement of the diffraction pattern, the structure parameters were constrained to correspond to the average I4/mmm model except for the Fe occupancies. The refined stoichiometry corresponds to $K_{1.06}Fe_{1.57}Se_2$. The lattice constants are $a$=3.94422 (28), and $c$=13.9306(18) Å, which are different from the values typical for samples obtained by high temperature synthesis: $a$=3.9092(2), $c$=14.1353(13) Å [2].

The diffraction pattern of the sodium intercalated sample shows a small number of Bragg peaks. The first peak could be indexed as (002) at 2Θ about 11°. Assuming that this peak originates from a small amount of a crystalline phase of $Na_x(C_5H_5N)_yFe_{2-z}Se_2$, the lattice constant $c$ can be estimated as $c$=13.7Å, which is also different from already reported value for the Na-intercalated sample (17.432(1) Å) [25].

The diffraction pattern of the rubidium intercalated sample shows the 122 phase similar to the potassium-intercalated one and 4% impurity of FeSe. The lattice constants are $a$=3.99035(37),

$c=14.2965(22)$ Å, which are close to the lattice parameters for the samples obtained by high temperature synthesis [6].

The diffraction pattern of $Li_x(C_5H_5N)_yFe_{2-z}Se_2$ presented in Fig.2 is significantly different from the K- and Rb- intercalated samples. There is a main phase identified with the metric corresponding to the 122-structure and two impurity phases: FeSe and $Fe_7Se_8$. In addition, there are three relatively small but visible Bragg peaks located at approximately $2\Theta=16.49$, 24.77, 27.15°. Tentatively we could assign these additional peaks to lithium cyanide Li(CN), however we are not sure in this model. The significant difference in comparison with the K-intercalated sample is a presence of a large anisotropic peak broadening. One can see that the peaks (101) at $2\Theta=25.68°$ and (114) at 42.36° are very broad. To describe the line broadening the anisotropic microstrain (dispersion of lattice constants) model in Fullprof was used [32]. The refined average microstrain $\Delta d/d$ value amounted to 2%. The peak positions are very well fitted to the I4/mmm crystal metric with $a = 3.55879(95)$ and $c = 16.0549(4)$ Å. However, the intensities of the peaks are not very well described by the 122-structure model, contrary to the K-sample. If one fixes the structure with the 122-structure model, many Bragg peaks intensities [e.g. (101) and (114)] are underestimated. This implies the presence of additional scattering density in the unit cell. As it was proved further with thermal analysis (see Fig.3) the pyridine molecules enter the structure together with lithium ($Li_x(C_5H_5N)_y$). On a Fourier density map at the $z = 1/2$ slice obtained from XRD data we have also seen non-zero electronic density areas located between the Li cations. The lattice constant $c$ is significantly larger than in the potassium sample, additionally favoring the above suggestion. The large microstrain value obtained can be caused by the inhomogeneous distribution of the co-intercalated pyridine molecules in the crystallites. In addition, the lattice constants of the Li-intercalated sample differ from the reported values for $Li_x(NH_3)_yFe_2Se_2$ obtained by the ammonothermal method (a=3.775(5), c=17.04(3) Å) [23]. Structural parameters for all samples are presented in Table 1.

**Table 1.** Room temperature structural parameters for the $A_x(C_5H_5N)_yFe_{2-z}Se_2$ pristine samples and the $Li_x(C_5H_5N)_yFe_{2-z}Se_2$ post-annealed sample (see text below) obtained from XRD. Space group I4/mmm, Fe in the (4d) position (0, 0.5, 0.25); Se in the (4e) position (0, 0, $z$), K/Na/Rb/Li in the (2a) position (0, 0, 0). An average elemental content obtained by micro XRF mapping is also listed in the table.

|  | K | Na | Rb | Li | Li annealed P4 |
|---|---|---|---|---|---|
| $a$ (Å) | 3.94422 (28) |  | 3.99035(37) | 3.55879(95) | 8.00283 |

| $c$ (Å) | 13.9306(17) | 13.6884(64) | 14.2965 (22) | 16.0549(4) | 23.09648 |
| --- | --- | --- | --- | --- | --- |
| Refined occupancy | $K_{1.06(1)}Fe_{1.57(2)}Se_2$ | | $Rb_{0.97(1)}Fe_{1.49(2)}Se_2$ | | |
| average elemental content (microXRF) | $K_{1.15(2)}Fe_{2.02(3)}Se_2$ | $Na_{1.23(9)}Fe_{1.85(6)}Se_2$ | $Rb_{1.21(3)}Fe_{1.85(2)}Se_2$ | not measured | |

We present in Fig.3 thermal analysis measurements for $Li_x(C_5H_5N)_yFe_{2-z}Se_2$ along with the mass spectrometry (MS) signals for m/e=52 and m/e=79. The MS signals correspond to the evolution of $C_4H_4^+$ resulting from pyridine ion fragmentation ($C_6H_5N^+ \rightarrow C_4H_4^+$ + HCN [33]) and a neat pyridine, respectively. Temperature induced de-intercalation of pyridine runs in three steps at different temperature intervals. During the first step at temperatures below 140°C, only ca. 0.2% mass loss is observed. This may be related to the residual amount of the free solvent present on the surface of the grains. After the second step, between 140 - 180°C, the mass loss reaches ca. 0.5%. Further heating causes total de-intercalation of the guest molecules which finishes at 320°C. The fact that the second and the third steps are observed at quite high temperatures and the highest amount of pyridine is released at about 320°C might be the indication of the presence of stable 4,4'-bipyridyl in the material or formation of stable bipyridynium complexes with residual alkali metal cyanide. However, we did not observe any additional peaks in the mass spectrum indicating the fragmentation of 4,4'-bipyridyl. We noted that the amount of the co-intercalated solvent depends on the type of alkali metal. It was found to be the highest in Li-containing samples, and increased from 1.85(5) mass.% to 5.6(2) mass.% with the time of intercalation. These amounts correspond to 0.068 and 0.208 mole of pyridine per formula unit, respectively. In the case of sodium and potassium doped samples, the amount of the co-intercalated solvent was comparable and reached 0.022 and 0.021 mole of $C_6H_5N$ for Na and K samples, respectively. In the case of the Rb-sample, about 0.15 mole of pyridine was incorporated into the host matrix. Taking into account the presence of quite stable bipyridynium complexes and the disordered character of the main phase of the as prepared material, we decided to anneal the Li-intercalated sample at 215°C in sealed quartz ampoule for over 50 hours in the hope of improving its homogeneity.

The magnetic susceptibility measurements were performed on the as prepared and annealed samples to confirm the superconducting properties of $Li_x(C_5H_5N)_yFe_{2-z}Se_2$. The temperature dependencies of the real part of the ac magnetic susceptibility (measured with a field amplitude $H_{ac}$= 1 Oe) for the $Li_x(C_5H_5N)_yFe_{2-z}Se_2$ samples are presented in Fig. 4. The upper curve corresponds to $Li_x(C_5H_5N)_yFe_{2-z}Se_2$ pristine sample and the lower one to the

$Li_x(C_5H_5N)_yFe_{2-z}Se_2$ sample, which was post-annealed at 215 °C. One can see that both curves for the Li-pristine and the post-annealed sample have a diamagnetic transition above 40K, but the improvement of the shielding fraction in the Li-annealed sample is drastic. This can be a consequence of a rearrangement and more homogenous distribution of the intercalated guests. The crystal structure of the annealed sample also underwent significant changes (see Table 1). The crystal metric can no longer be described by I4/mmm space group with the lattice constants similar to the original sample. The first peak (002) at $2\Theta \approx 11°$ seems to move to $2\Theta \approx 7.5°$ implying a very large change in the c-constant reaching $c = 23$ Å. Instead of broad diffraction peaks, the annealed sample has many narrow peaks (Fig. 5). We were able to index most of the peaks using the tetragonal metric shown in the Table 1. The *a*-parameter is roughly √5 times larger, and the *c*-parameter was enlarged from 16 to 23 Å, which can be explained, for example, by rotation of the aromatic rings in bi-pyridyl moieties incorporated between the host layers (FeSe). We attempted to use different smaller cells (for example simply using I4/mmm with similar *a*-constant and enlarged *c*-constant), but without any a conclusive result. We therefore do not have any definite structural model, but the 122 structure model does not fit the peak intensities either, contrary to [23].

The insert of the Fig.4 shows the real part of the ac susceptibility for the $A_x(C_5H_5N)_yFe_{2-z}Se_2$ (A=Na, Rb and K). A decrease in the respective magnetic susceptibilities with decreasing temperature around 35K, 10K and 15K for sodium, potassium and rubidium doped samples, respectively, may hint to spurious superconductivity also in these compounds.

The temperature dependence of the resistivity of a polycrystalline sample of $Li_x(C_5H_5N)_yFe_{2-z}Se_2$ is shown in Figure 6. The resistivity in the normal state shows above 100 K a weakly metallic behavior. Below 100 K, the resistivity increases and shows a semi-conducting behavior above 45 K. In the resistivity measurement, a $T_{C,onset}$ of 45 K for $Li_x(C_5H_5N)_yFe_{2-z}Se_2$ and at 31 K a second transition (here denoted as $T_{C2}$) was observed. The impurity phase FeSe cannot be seen in the resistivity measurement because the sample reaches a state of zero resistance (10 K) above the critical temperature of FeSe ($T_C \sim 8$ K). In the inset of Figure 6, we show the field dependence of transition to superconductivity. These measurements are a clear indication that $Li_x(C_5H_5N)_yFe_{2-z}Se_2$ is a new bulk superconductor.

## 4. Summary

In the present work we report on a new room temperature synthesis route for 122 iron selenide superconductors $A_x(C_5H_5N)_yFe_{2-z}Se_2$ (A = Li, Na, K, Rb) by intercalation of alkali metals from an organic solvent into the FeSe matrix. The Li-containing sample clearly shows a superconducting transition at $T_C \sim 45K$ as shown by the resistivity measurements. Post-treatment of the Li-intercalated sample at elevated temperatures drastically changes the c-parameter of the unit cell and increases the amount of the superconducting shielding fraction, which can be a consequence of a rearrangement and a more homogenous distribution of the intercalated alkali metals and organic solvent.

**Acknowledgements**

The work at the Universität Zürich was partly supported by Forschungskredit UZH. The authors would like to thank Casey Marjerrison for assisting with the preparation of the manuscript.

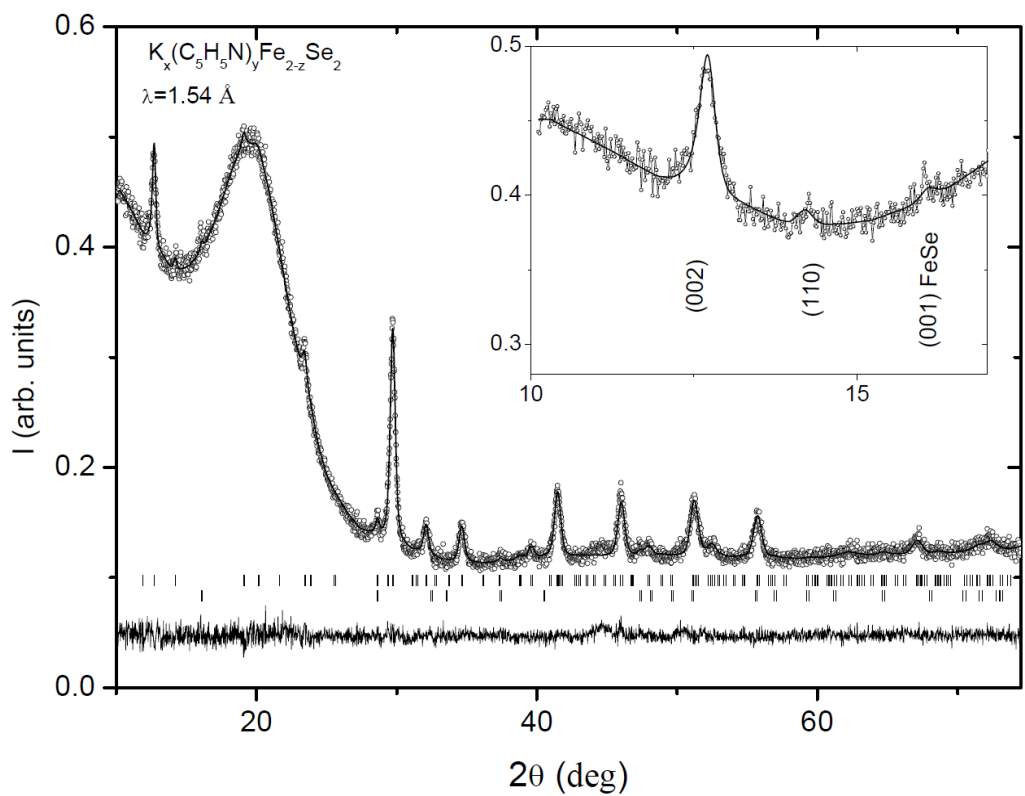

**Figure 1.** (a) X-ray powder diffraction pattern, the calculated profile and difference plot for $K_x(C_5H_5N)_yFe_{2-z}Se_2$ at 300 K. The rows of ticks show the Bragg peak positions for the 122 and 11 phases. (b) Fragment of the diffraction pattern at low $2\Theta$ showing superstructure satellite (110). Its intensity corresponds to completely unoccupied Fe-vacancy site.

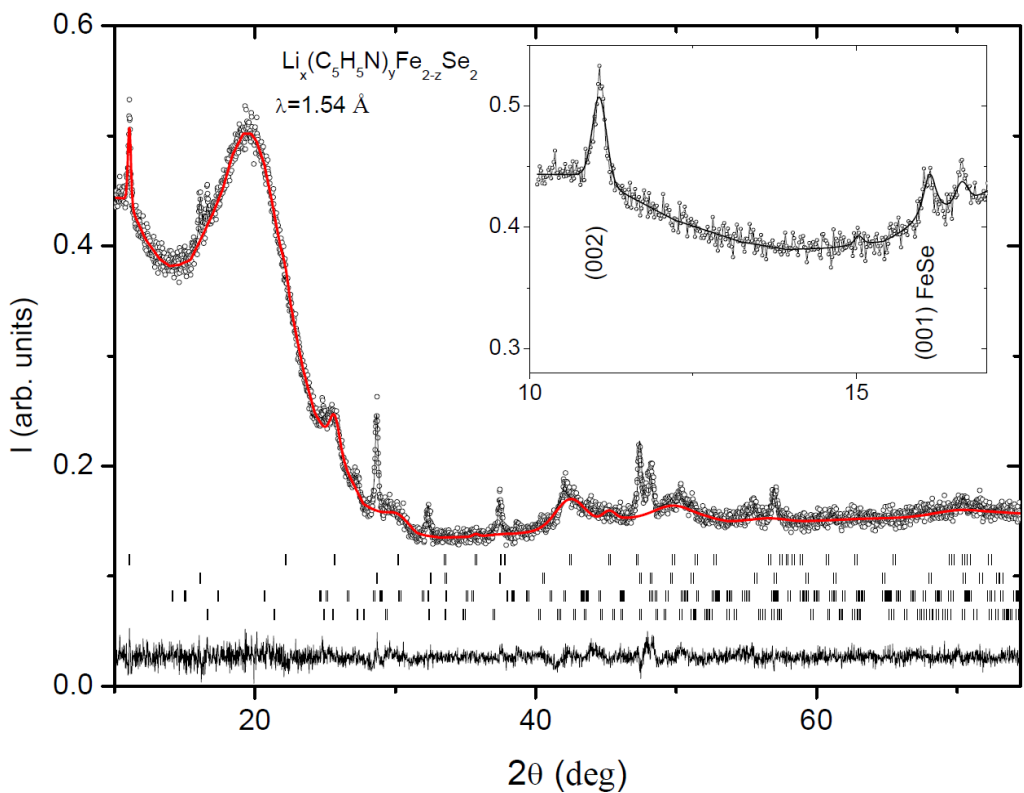

**Figure 2.** (a) X-ray powder diffraction pattern, the calculated profile and difference plot for $Li_x(C_5H_5N)_yFe_{2-z}Se_2$ at 300 K. The rows of ticks show the Bragg peak positions for the 122, 11 FeSe and $Fe_7Se_8$ phases, and the red line represents Lebail mode refinement without structural fitting. (b) Fragment of the diffraction pattern at low 2Θ showing (002) reflection of 122 phase and (001) belonging to 11 FeSe.

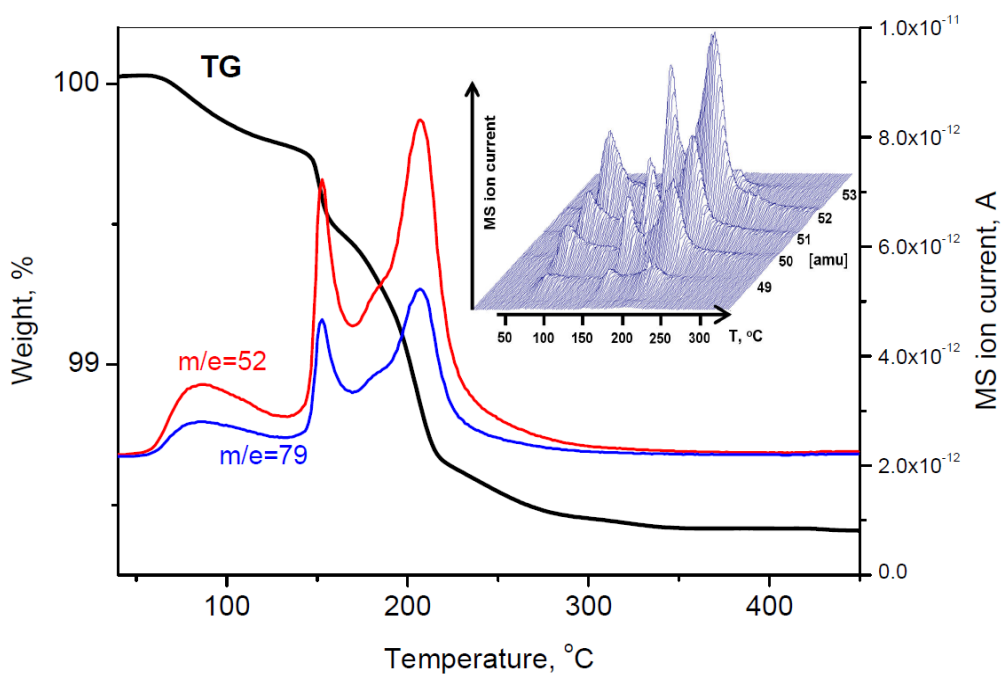

**Figure 3.** Thermogravimetric curve for $Li_x(C_5H_5N)_yFe_{2-z}Se_2$ sample together with signals from the mass-spectrometer for the ions m/e=52 and m/e=79. The insert shows the temperature evolution of m/e=49-53 signals corresponding to different fragmentation of $C_4H_4^+$.

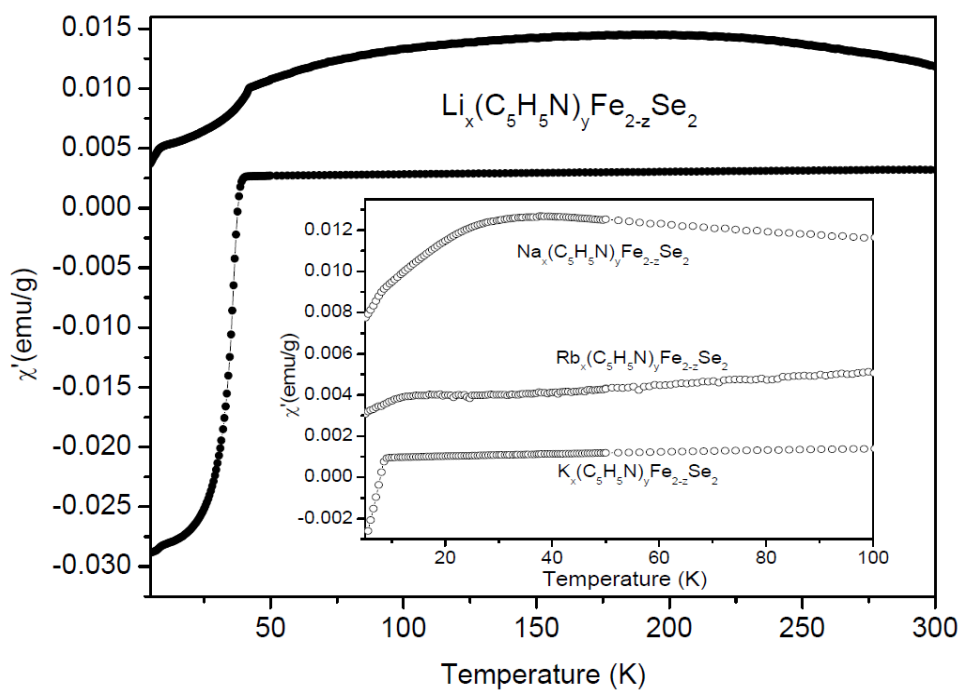

**Figure 4.** Temperature dependence of the real part of the ac magnetic susceptibility for $A_x(C_5H_5N)_yFe_{2-z}Se_2$ samples. The upper curve corresponds to a $Li_x(C_5H_5N)_yFe_{2-z}Se_2$ pristine sample, the lower one to $Li_x(C_5H_5N)_yFe_2Se_2$ sample post-annealed at 215 °C. In the insert the real part of the ac magnetic susceptibilities for $A_x(C_5H_5N)_yFe_{2-z}Se_2$ (A=Na, Rb and K) samples are presented.

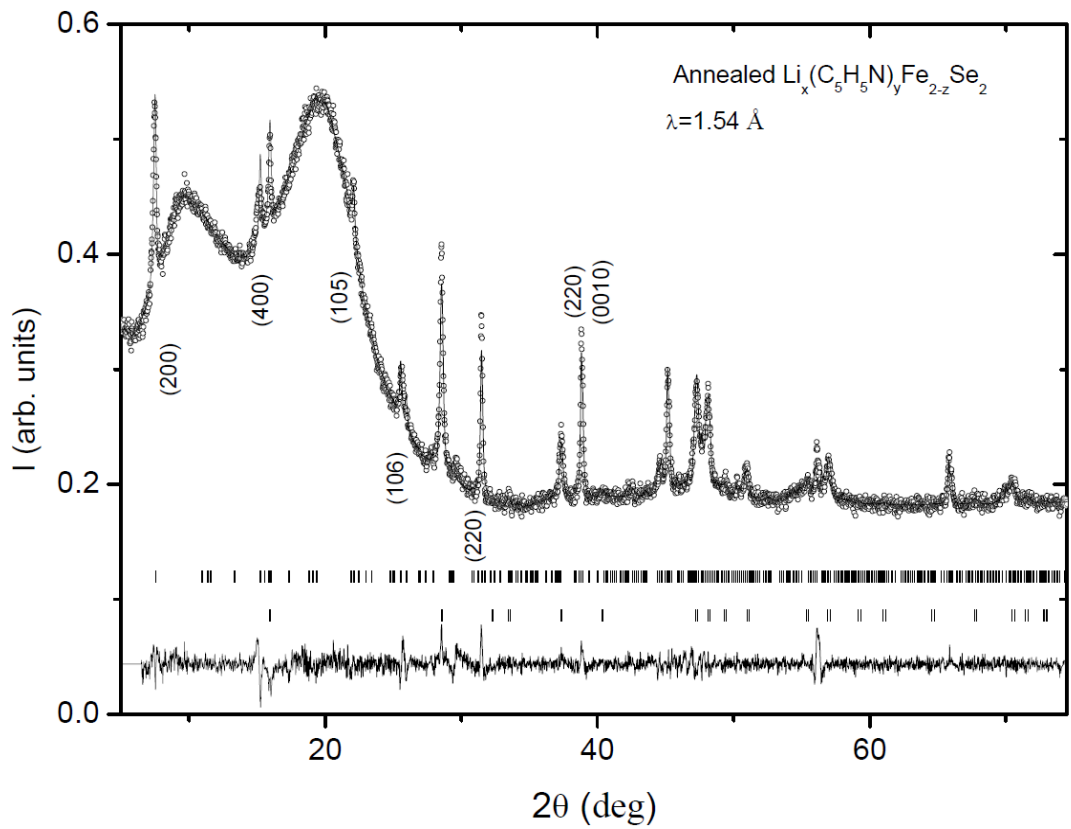

**Figure 5.** (a) X-ray powder diffraction pattern, the calculated profile and difference plot for the annealed sample of $Li_x(C_5H_5N)_yFe_{2-z}Se_2$. The main phase was refined in Lebeil mode in tetragonal cell with $a$ = 8.00283 and $c$ = 23.09648 Å. The rows of ticks show the Bragg peak positions for the tetragonal P4 main phase and 11 FeSe. The first diffraction peaks of the main phase are labeled.

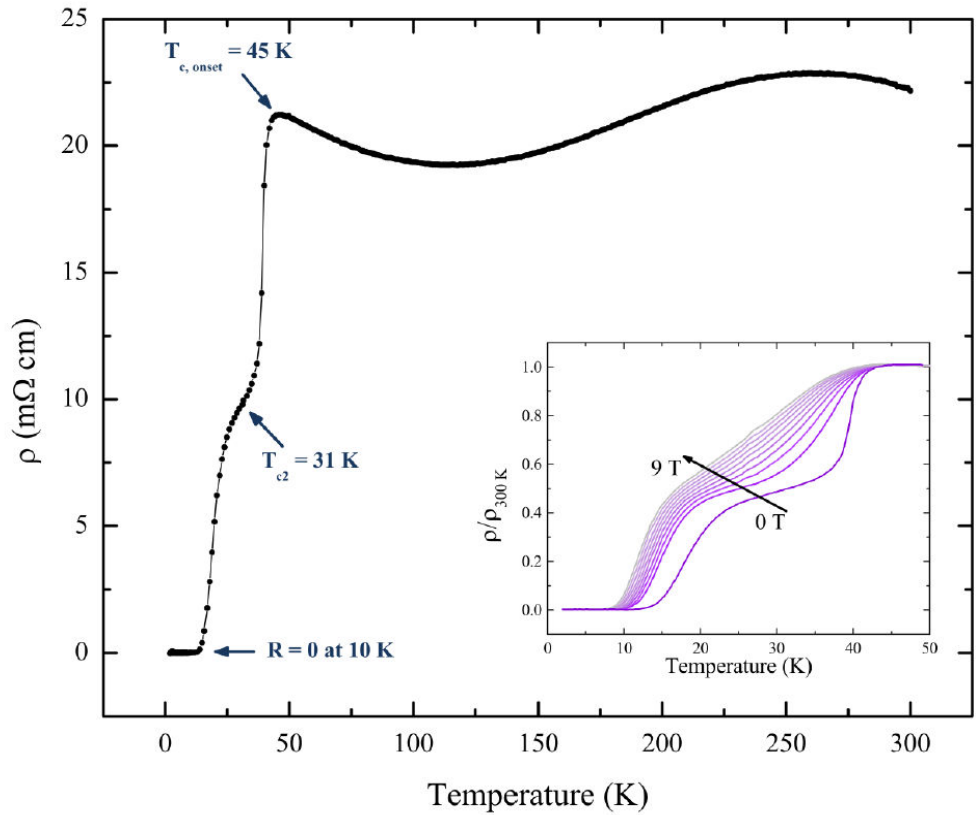

**Figure 6.** Temperature dependence of the electrical resistivity ρ of the as-prepared $Li_x(C_5H_5N)_yFe_{2-z}Se_2$ sample. The insert shows magnetic field dependence of the resistivity.